\documentstyle[11pt,newpasp,twoside]{article}
\def\edcomment#1{\iffalse\marginpar{\raggedright\sl#1\/}\else\relax\fi}
\reversemarginpar

\begin{document}
\title{Tracing Cosmic Evolution with the Las Campanas Distant Cluster Survey}
 \author{Anthony H. Gonzalez}
\affil{Harvard-Smithsonian Center for Astrophysics, 60 Garden Street, 
Cambridge, MA 02138, USA}
\author{Julianne J. Dalcanton}
\affil{Department of Astronomy, University of Washington, Box 351580, 
Seattle, WA 98195, USA}
\author{Amy E. Nelson}
\affil{Department of Astronomy and Astrophysics, University of California, 
Santa Cruz, CA 95064, USA}
\author{Luc Simard \& Dennis Zaritsky}
\affil{Steward Observatory, University of Arizona, 933 North Cherry Avenue,
Tuscon, AZ 85721, USA} 
\author{Risa H. Wechsler}
\affil{Department of Physics, University of California, Santa Cruz, CA 95064}

\begin{abstract}
The Las Campanas Distant Cluster Survey (LCDCS), which contains over 1000
cluster candidates at $z$$>$0.3, is a unique sample with which to
probe the evolution of both cluster galaxies and the properties of the 
cluster population. Programs are now underway to utilize the LCDCS for
both purposes. We provide a brief overview of these programs, and also
present new results for the LCDCS cluster correlation function. Utilizing
well-defined, approximately dispersion-limited subsamples, we measure the
angular correlation function for clusters at $z$$\approx$0.5. Spatial correlation
lengths are then derived via Limber inversion. We find that the correlation 
length depends upon mass, as parameterized by the mean cluster separation,
in a manner that is consistent with both local observations and CDM 
predictions for the clustering strength at $z$$=$0.5.
\end{abstract}

\section{Summary}

The past few years have been a period of tremendous progress in detection
of distant clusters. As recently as five years ago no single survey contained
more than ten cluster candidates at $z$$>$0.5, but a series of efforts have
since led to a several order of magnitude increase in the number of known 
high-redshift systems (Postman et al. 1996, Scodeggio et al. 1999, Gladders 
\& Yee 2001, Gonzalez et al. 2001).
The recently completed Las Campanas Distant Cluster
Survey (LCDCS) is our contribution to this field.
Clusters in the LCDCS are detected as regions of excess optical surface
brightness relative to the background sky, a technique that permits detection
of clusters to $z$$\approx$0.9 with shallow, drift-scan imaging. The resulting
statistical catalog consists of 1073 cluster candidates at $z$$\ga$0.3.
To enable efficient use of these catalogs we use
follow-up imaging and spectroscopy to characterize the properties of the
sample, including the contamination rate as well as estimates of the redshifts
and relative velocity dispersions of the candidates (see Gonzalez et al. 2001$a$ 
for details).

One application of this sample is determination of the cluster
correlation function (Gonzalez et al. 2001$b$). For roughly
dispersion-limited subsamples at $z$$\approx$0.5 the observed angular correlation
functions are well fit by the power law
$\omega(\theta)=A_\omega \theta^{1-\gamma}$ with $\gamma=2.1$.
Using the angular correlation data, we derive the spatial
correlation length ($r_0$) as a function of mean intracluster separation ($d_c$)
via the cosmological Limber inversion. We find correlation lengths for the 
LCDCS that are similar to those observed for local catalogs, and observe a 
dependence of $r_0$ upon $d_c$ that is comparable to the results of Croft et
al. (1997) for the APM catalog. These results are consistent with theoretical
expectations, which predict only modest evolution in the clustering amplitude
since $z$=0.5. 

Several programs are also being conducted with LCDCS clusters  to study the 
evolution of cluster galaxies.
Nelson et al. (2001$a,b$) are studying the structural and luminosity evolution 
of BCG's; Nelson et al. (2001$c$) are studying evolution in the luminosity
function and color of the red envelope. Meanwhile, the EDisCS collaboration is
conducting a deep multicolor imaging and spectroscopic survey of 20 of the most
massive LCDCS candidates at $z$=0.5 and $z$=0.8, with the aim of constraining
cluster galaxy morphologies, scaling relations, cluster dynamics, and
the evolution of cluster galaxies at these epochs.

\end{document}